\documentclass[5p,times,twocolumn]{elsarticle}

\usepackage{lineno,hyperref}
\usepackage[utf8]{inputenc}
\modulolinenumbers[5]

\begin{document}

\begin{frontmatter}

\title{Series Testing and Characterization of 1100 Hamamatsu H12700 Multianode Photomultiplier Tubes}

\author[b]{J.~F\"ortsch\corref{mycorrespondingauthor}}
\ead{foertsch@uni-wuppertal.de}
\author[b]{K.-H.~Kampert}
\author[b]{V.~Patel}
\author[b]{C.~Pauly}
\author[b]{D.~Pfeifer}

\address[b]{Department of Physics, University of Wuppertal, D-42097 Wuppertal, Germany}
\cortext[mycorrespondingauthor]{Corresponding author.}

\begin{abstract}
Both the future CBM-RICH and the recently upgraded HADES-RICH use Hamamatsu H12700 Multi-Anode Photomultipliers (MAPMTs) as their photon detectors.
To test the MAPMTs thoroughly before using them in the detectors, the photon sensors were qualified well in advance and a test bench was built to efficiently characterize each MAPMT.
The test bench measures the single photoelectron gain, dark rate, relative efficiency, and afterpulse probability of each channel.
This article describes the operating principle of the test bench and discusses the distributions of each measured quantity over 1,100 MAPMTs.
Additionally, the long-term aging effects of the H12700 MAPMT are investigated based on repeated measurements of individual MAPMTs.
\end{abstract}


\date{August 2025}
\end{frontmatter}


\section{Introduction} 
\label{sec:introduction}
The Hamamatsu H12700 MultiAnode Photomultiplier tube (MAPMT) is a $52\times52\,\mathrm{mm}^2$ \footnote{First MAPMTs of Type H12700 had a form factor of $52\times52\,\mathrm{mm}^2$, whereas newer models feature a form factor of $51.7\times51.7\,\mathrm{mm}^2$.} MAPMT with $8\times8$~individual amplification channels\cite{Datasheet} of size $6\times6\,\mathrm{mm}^2$. The tube is read out via 64~separated anode pads (pixels) marking the end of each dynode chain, realized in a metal channel configuration.
The sensor is well-suited for Cherenkov applications, featuring a bialkali photocathode with a quantum efficiency of $\ge 30\%$, peaking at a wavelength of $350\,\mathrm{nm}$ \cite{Calvi:2015yra} and extending into the UV region due to its UV-transparent front glass.
Since most Cherenkov detectors require the ability to resolve single photons, this sensor must have a low dark rate and a well-separable single photoelectron peak in each amplification channel.

The CBM experiment aims to measure the QCD phase diagram at high baryon densities using high-energy heavy-ion collisions at the future FAIR facility at GSI in Darmstadt, Germany.
An essential part of the CBM detector setup is  the Ring Imaging Cherenkov detector (RICH)~\cite{rich_proj}, used for particle identification, in particular electron/pion separation.
The CBM-RICH uses CO$_2$ as radiator gas to produce Cherenkov light. This light is guided via a spherical mirror (split into two individually tilted half mirrors) onto two photon detection planes, where rings form.
Each of the two photon detection planes will be equipped with up to 588 H12700 MAPMTs, which will be read out by DiRICH readout modules, as described in Ref.~\cite{Michel_2017}.
The same MAPMTs, along with their readout chains are incorporated into the upgraded HADES-RICH detector.
The HADES-RICH detector is part of the HADES spectrometer, which is currently in operation at GSI.
The HADES experiment aims to measure the properties of matter at medium temperatures and high densities through heavy-ion collisions.
The upgraded HADES RICH detector was successfully used for the first time during a four-week Ag+Ag production beamtime in 2019. It uses \textit{i}-C$_{4}$H$_{10}$ as radiator gas for electron identification. The produced Cherenkov light is reflected by an imaging spherical mirror onto 428~MAPMTs arranged in a circular footprint.

To equip the HADES-RICH and CBM-RICH with H12700 MAPMTs, 1,100 of these MAPMTs were ordered from Hamamatsu.
A dedicated test bench was built to provide immediate feedback during the production and delivery of the MAPMTs, as well as to fully characterize each MAPMT, e.g. in terms of gain for further grouping.
The measurements and results obtained from this test bench are discussed in the following.


\section{The test bench setup} 
\label{sec:the_test_bench_set_up}
One of the main requirements for designing the test bench was to fully characterize all MAPMTs down to the pixels level, soon after delivery.
With 50 MAPMTs delivered each month, the test bench was constructed to allow scanning three MAPMTs (plus a fourth MAPMT used as a reference) simultaneously within eight hours.
The following characteristics were emphasized:
\begin{itemize}
    \item Single photoelectron gain per pixel
    \item Gain-voltage dependence
    \item (Spatially resolved) relative efficiency measurement
    \item Single pixel dark rate 
    \item Average afterpulse probability
\end{itemize}

\begin{figure}[tb]
	\centering
	\includegraphics[width=\linewidth]{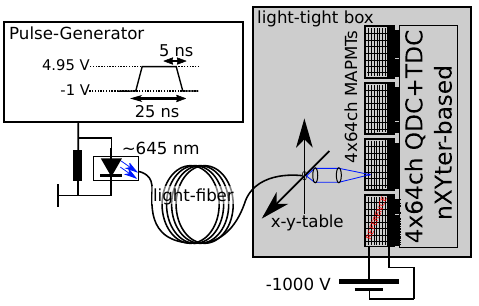}
	\caption{Sketch of the test bench also showing the configuration of the pulse generator, variable supply voltage (HV) and light source.}
	\label{fig:test_bench}
\end{figure}

\begin{figure}[tb]
	\centering
	\includegraphics[width=0.7\linewidth]{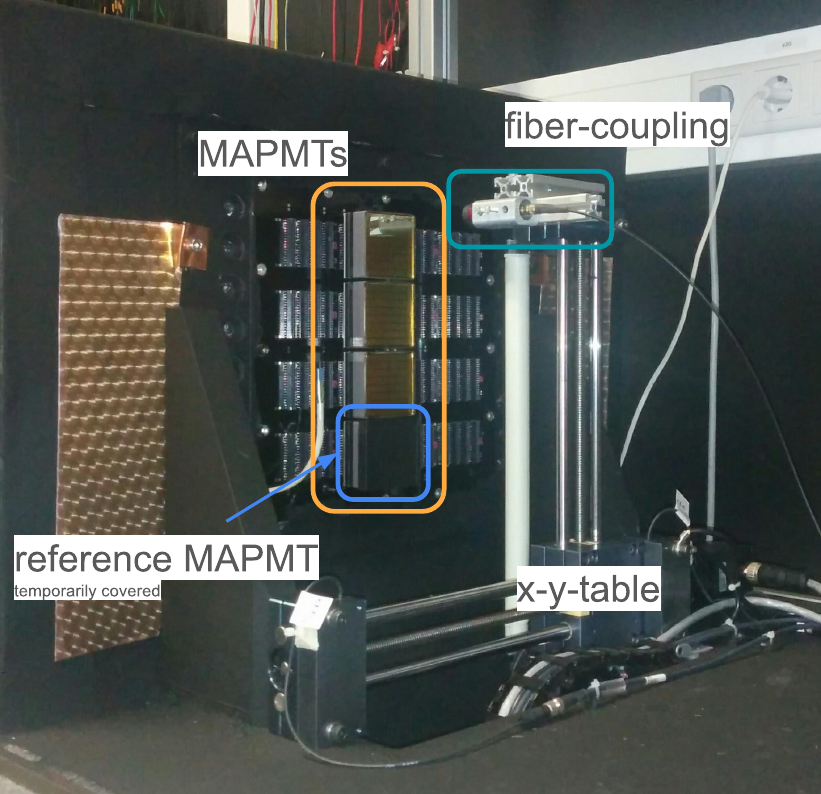}
	\caption{Picture of the test bench with opened light tight box but covered reference MAPMT. The MAPMTs, the light guide, the reference MAPMT and the $x$-$y$-table are indicated with rectangles.}
	\label{fig:test_bench_pic}
\end{figure}

To measure the relative efficiency and single photoelectron gain, a light fiber attached to an $x$-$y$-table was used as a ``single photon'' light source. The far end of the fiber was coupled to a pulsed LED with a wavelength of $465\,\mathrm{nm}$ and the following settings: $25\,\mathrm{ns}$ width, $5\,\mathrm{ns}$ slope, $-1\,\mathrm{V}$ ground, $4.950\,\mathrm{V}$ peak, $13\,\mathrm{kHz}$ pulse-frequency. Dimming was achieved by increasing the distance to $20\,\mathrm{cm}$ in the fiber coupling between the LED and the fiber.
The light level at the output was chosen such that, on average, every tenth pulse would lead to a measured signal at the MAPMT. This low intensity ensures a double-photon probability within the same pulse of only $\sim$$0.5\%$. A fiber focusing optic was used to induce a spot size of around $1.5\,\textrm{mm}$ on the MAPMT.
Using the $x$-$y$-table, the surface of 3+1 MAPMTs mounted in the setup are scanned in steps of $(1.000\pm0.002)\,\mathrm{mm}$.
The fourth MAPMT serves as the reference MAPMT and remains in the measurement setup for each measurement.
This allows for a relative efficiency comparison of all measured MAPMTs, independent of long-term light intensity variations.
Cross-checks were performed at regular intervals by remeasuring the MAPMTs from the first delivery charge (stored in a light-tight cabinet flushed with N$_2$) to test for aging effects on the regularly used reference MAPMT.
At each of the $55\times216$ measurement points, the $x$-$y$-table stops for $1.5\,$s to gather $\sim$$2000$ single photon signals.
To ensure proper measurement of dark rate and afterpulse probability, all four MAPMTs are enclosed in a light-tight setup.
Figure~\ref{fig:test_bench} shows a sketch of the setup, and Fig.~\ref{fig:test_bench_pic} a photo of it, where the vertical arrangement of the four MAPMTs, each normally supplied with $-1000\,\mathrm{V}$, is visible.
A complete measurement sequence begins with a one-hour ``cool-down'' period, during which the MAPMT is supplied with $-1\,$kV and kept in darkness.
This is followed by the main measurement scan described above. In a final step, a measurement dependent on the supply voltage is carried out in which the MAPMT is operated sequentially at six different supply voltages, ranging from $-850\,$V to $-1100\,$V. For each voltage setting, a subset of pixels\footnote{The pixels 64, 55, 46, 37, 28, 19, 10, 1, 57, 50, 43, 36, 29, 22, 15 and 7 are measured with slightly extended illumination times to ensure that approximately 6500 photon signals are recorded per pixel.} is measured at their corresponding center positions to extract the key parameters for the given supply voltage.


The free streaming n-XYter-FEB (cf.~\cite{nxyter}) is used for the readout of each MAPMT pixel, measuring the charge (QDC) and arrival time (TDC) of each pulse.
Using the self-triggered, multi-hit capable n-XYter electronics for parallel readout of all $4\times64$~MAPMT pixels enables us to determine various characteristic quantities of the MAPMTs from a single scan dataset, as will be explained in the following paragraphs.\\

By correlating each measured pulse with the trigger that initiated the LED pulse, one can distinguish between different types of signals:
A signal measured within a time window of $\pm50\,\mathrm{ns}$ around the average LED photon arrival time is assumed to be directly correlated with the LED pulse.
Signals measured before this window (from $-1925\,\mathrm{ns}$ to $-325\,\mathrm{ns}$) are assumed to be uncorrelated with the LED light source and are attributed to the MAPMT pixel dark rate.
Signals in the time window after the LED signal (from $50\,\mathrm{ns}$ to $3075\,\mathrm{ns}$) consist of afterpulses (with a delay of more than $50\,$ns, Type 2, as described on \cite[p.~79]{HAMPhotonBook}) as well as dark rate signals.
Subtracting the dark rate (measured in the time window before the trigger) separates these two contributions, enabling statistical statements regarding the afterpulse probability.

\noindent {\bf Spatially resolved relative photon detection efficiency:}
To determine the spatially resolved relative efficiency at each measurement point, the number of single photon pulses detected during the $1.5\,\mathrm{s}$ measurement period is counted and divided by the number of trigger pulses measured during that time.
During analysis, this number is further normalized by the surface-averaged number of detected photon signals on the reference MAPMT in order to counteract long-term variations in the LED's intensity.
This method of deriving a relative efficiency was established in \cite{PAULY2022167177} where it delivers robust results when comparing irradiated photon sensors.
With this normalization, the efficiency of each measurement point on a given MAPMT can be viewed relative to the reference MAPMT, which is an MAPMT with an average efficiency.
To quantify each MAPMT's efficiency, the so-called ``Efficiency-Index'' is derived. It represents the spatial-averaged single photon efficiency of an individual MAPMT relative to the reference MAPMT.
The uncertainty in this average relative efficiency is estimated by taking repeated measurements of the same MAPMTs. For a common efficiency of $1.000$, the uncertainty is found to be $\pm0.006$.

\noindent {\bf Gain:}
As for the efficiency calculation, the single photoelectron gain per pixel is derived using the laser-correlated signals ($50\,$ns prompt time window).
The charge spectrum of the prompt signals is measured separately for each pixel, using only the signals measured when the light fiber is in the central $2\times2\,\textrm{mm}^2$ area of the respective pixel.
The single photoelectron gain is determined by fitting the primary peak in each charge spectrum with a Gaussian, where the mean of the fit corresponds to the gain.
After applying a charge calibration\footnote{The charge calibration was derived on single channels using well-defined charge pulses.} to the n-XYter channels, the fitted mean pulse charge is divided by the elementary charge and stored as the gain for each individual MAPMT pixel.
Through repeated measurements of the same MAPMT, the uncertainty on the measurement was determined to be $0.25\times10^6$ with a typical gain for the H12700 MAPMT of $\sim$$1.5\times10^6$ according to the data sheet (cf.~\cite{Datasheet}). 
For gains below approximately $1.0\times10^6$, the limited dynamic range of the ADC prevents peak fitting. In such cases, the peak position is estimated by extrapolating from the upper limit of the charge spectrum. At normal supply voltages of $-1\,$kV such low gains are never reached, so that this technique is only used to study supply voltage dependencies.

\noindent {\bf Dark rate:}
The dark rate of PMTs depends heavily on the previous illumination of the photocathode and it might take several hours of operation under low-light conditions before the dark rate reaches a stable value for each individual PMT.
Therefore, determination of the MAPMT dark rate per channel is only done using data from the end of the full scan set. This data is collected after a minimum of five hours, during which the MAPMT is supplied with high voltage and kept in darkness being only illuminated sporadically with single photons at low rates.
Furthermore, the MAPMT dark rate depends heavily on its temperature. Since the temperature of the measurement setup is not actively controlled and is therefore subject to daily weather changes, the dark rate must be corrected for this effect.
To accomplish this, the temperature is constantly measured during the entire measurement and the derived dark rate value is corrected using the formula outlined in \cite[pp.~74,75]{Selyuzhenkov:201318}.
With this correction, the absolute uncertainty of the average dark rate of a measured MAPMT was found to be in the range of $0.30\,$kHz with total dark rate values for all measured MAPMTs averaging $1\,$kHz as sum over all 64 pixels.

\noindent {\bf Afterpulsing:}
As described above, the determination of the afterpulse probability is based on data from a time window of $50\,\mathrm{ns}$ to $3075\,\mathrm{ns}$ after the LED trigger, conditioned on the presence of a correlated prompt signal in the same or a neighboring MAPMT pixel.
Due to limited per-channel statistics, the afterpulse analysis is performed as an average across all pixels of an MAPMT. 
By analyzing the fluctuations in the afterpulse window of the reference MAPMT over several measurements, it was determined that the uncertainty of this measurement amounts to $0.05\,$percentage points (pp.) for an average afterpulse probability of $0.93\%$.


\section{Average characteristics of 1100 MAPMTs} 
\label{sec:average_characteristics_of_1100_mapmts}

The acquisition of the largest data set to date, comprising 1,100 measured H12700 MAPMTs produced by Hamamatsu 
over a period of 3 years, 
enables a detailed study of the series spread in several key performance parameters.
This section analyzes variations in gain, dark rate, homogeneity in efficiency across the photocathode areas, and afterpulsing probability.

\subsection{Single photoelectron gain} 
\label{sub:single_photoelectron_gain}
\begin{figure}[hbt]
    \centering    \includegraphics[width=0.75\linewidth]{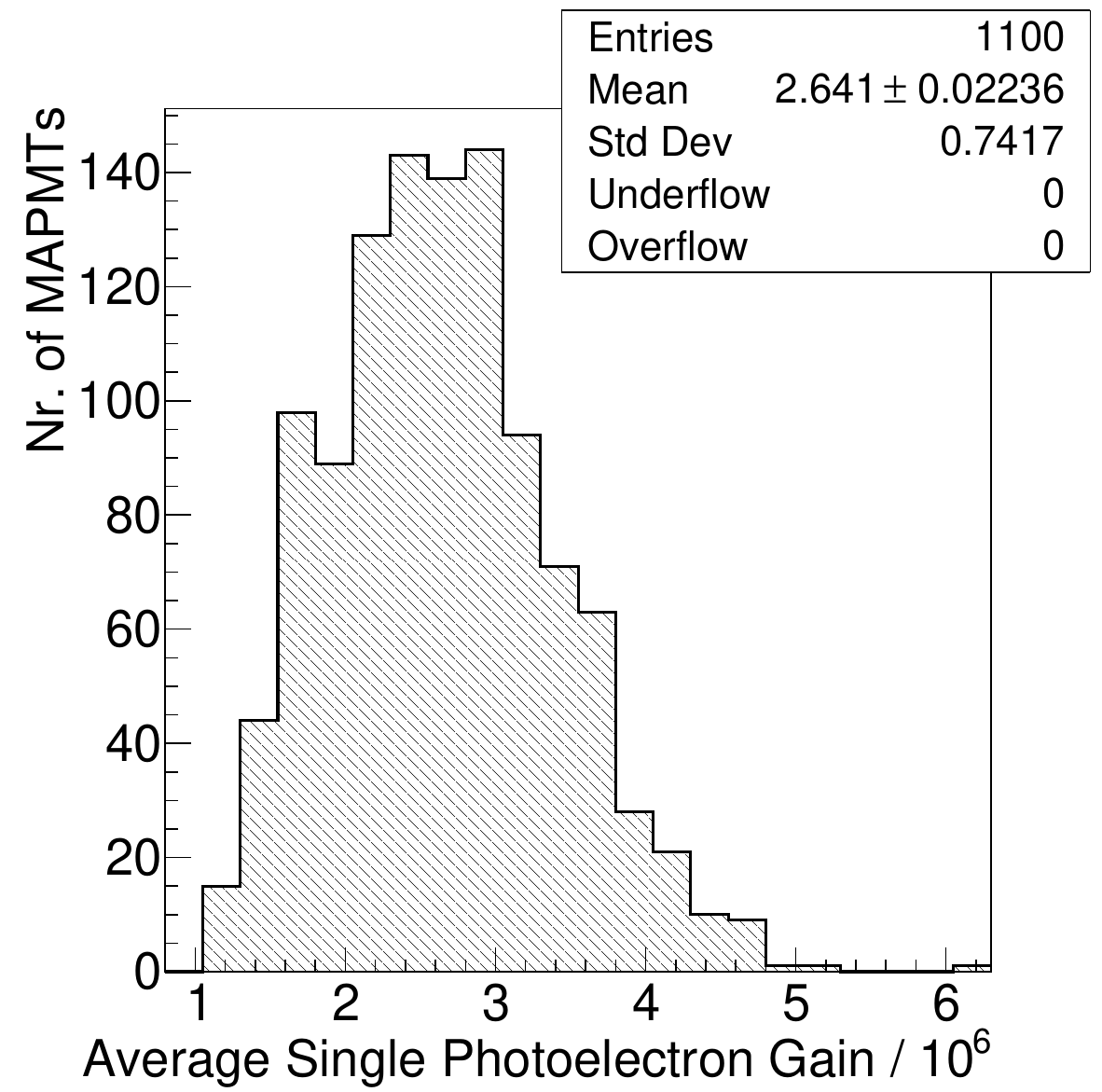}
	\caption{Single photoelectron gain at $-1000\,$V average of all 64 channels for each of the 1100 H12700 MAPMTs.}
    \label{fig:gain_hist}
\end{figure}

Figure~\ref{fig:gain_hist} shows the spread in single photoelectron (SPE) gain averaged over all 64 channels of a given MAPMT for a nominal supply voltage of $-1000\,$V.

\begin{figure}[htb]
    \centering\includegraphics[width=0.75\linewidth]{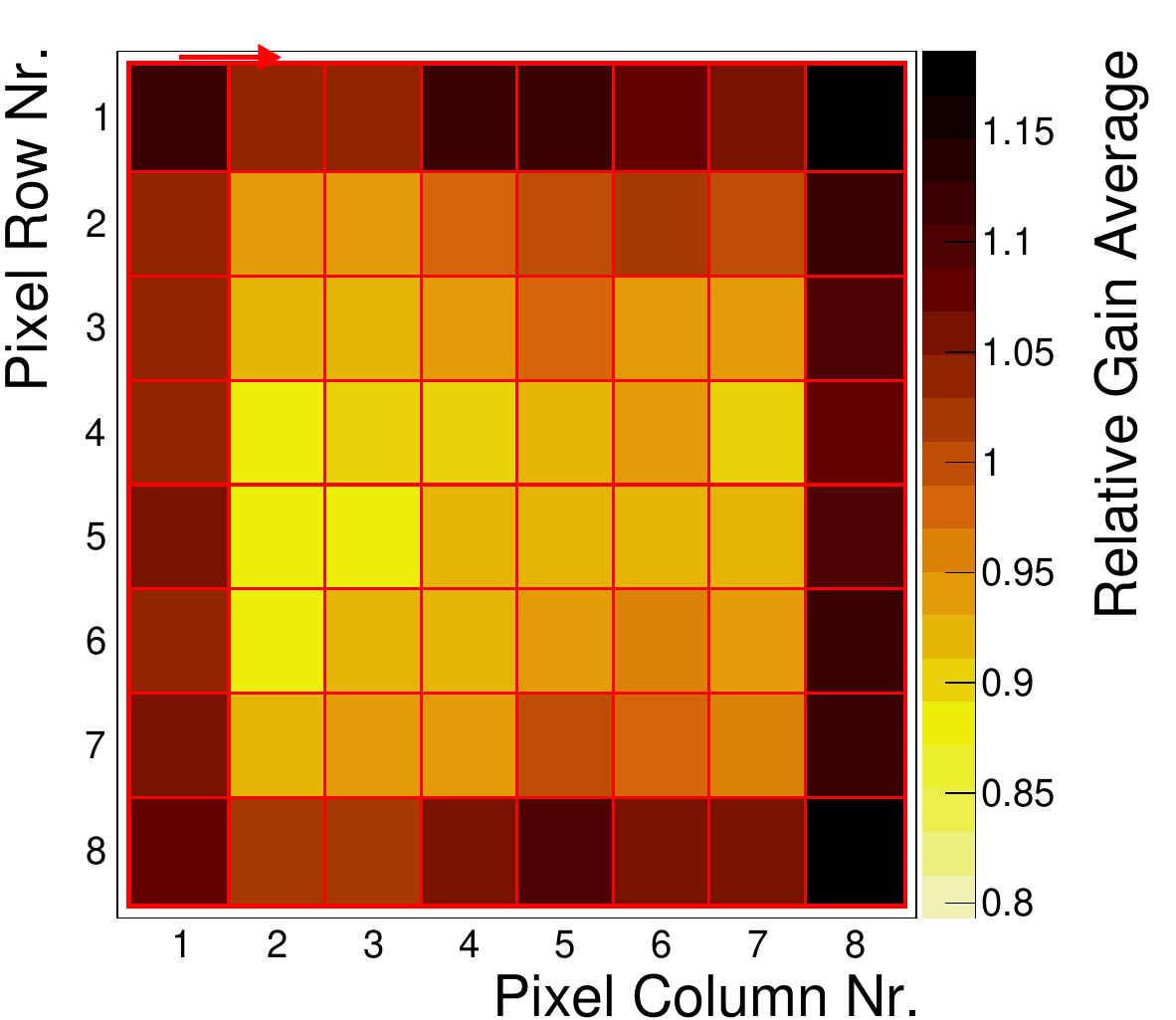}
	\caption{The relative variation of the pixel SPE gain is depicted in two dimensions, averaged over all MAPMTs and normalized to the average MAPMT SPE gain. The Hamamatsu ``START MARK'' is indicated by an arrow in the top left corner.}
    \label{fig:gain_var_hist_2d}
\end{figure}

\begin{figure}[htb]
	\centering	\includegraphics[width=0.869\linewidth]{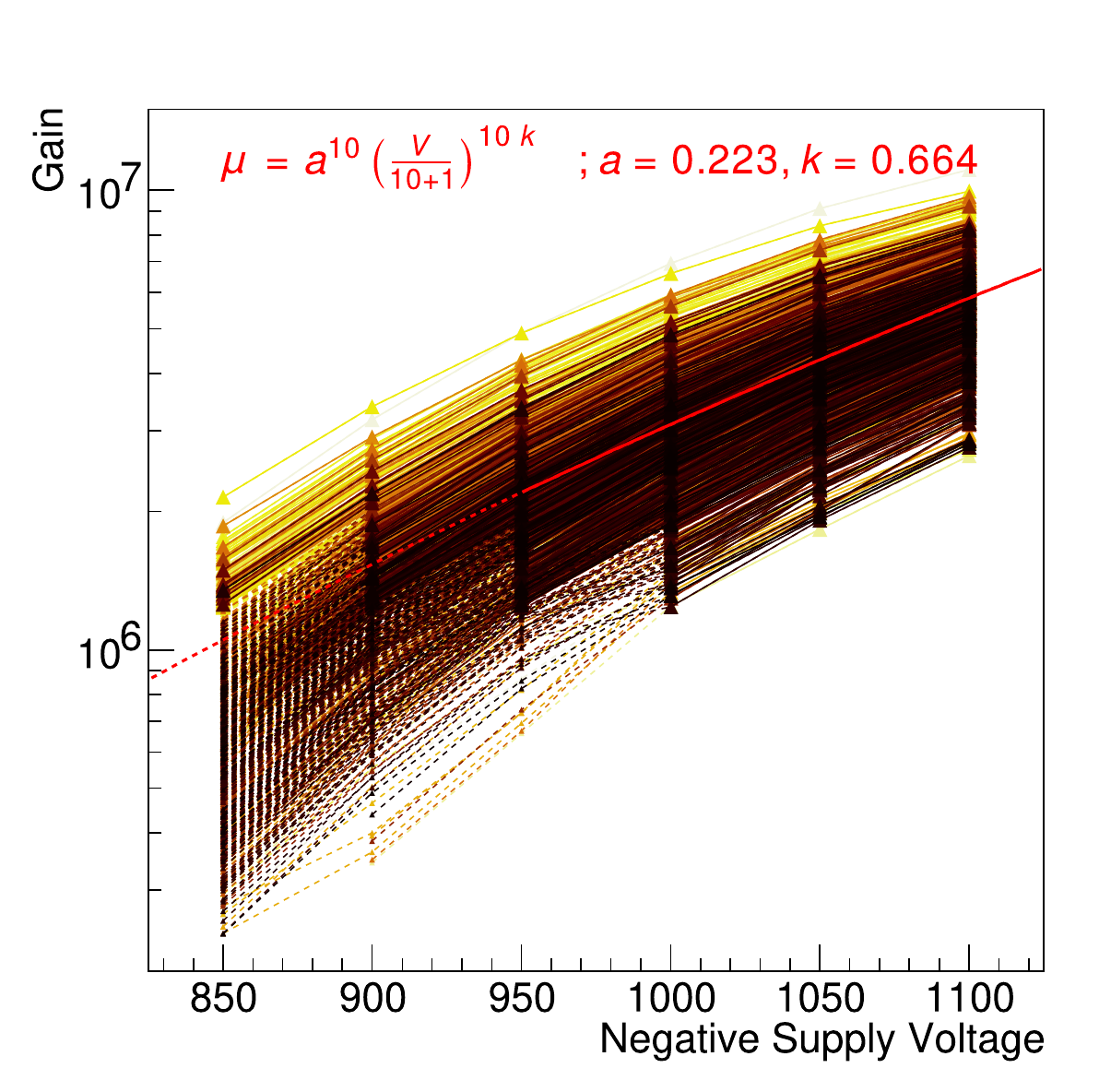}
	\caption{Single photoelectron gain as a function of supply voltage. Each marker shows the measurement for one MAPMT supplied with a particular voltage. The lines serve as guides and the color is chosen arbitrarily based on the MAPMT's serial number. Due to the limited dynamic range of the ADC, the measured gain values close to and below $1\times10^6$ are extrapolated (cf.\ section~\ref{sec:the_test_bench_set_up}) and are indicated by small markers and dashed lines. The red line shows a fit to Eq.~\ref{ham_eq}. The resulting parameters are displayed in the figure.}
	\label{fig:gain_vs_HV}
\end{figure}

\begin{figure}[htb]
	\centering	\includegraphics[width=0.75\linewidth]{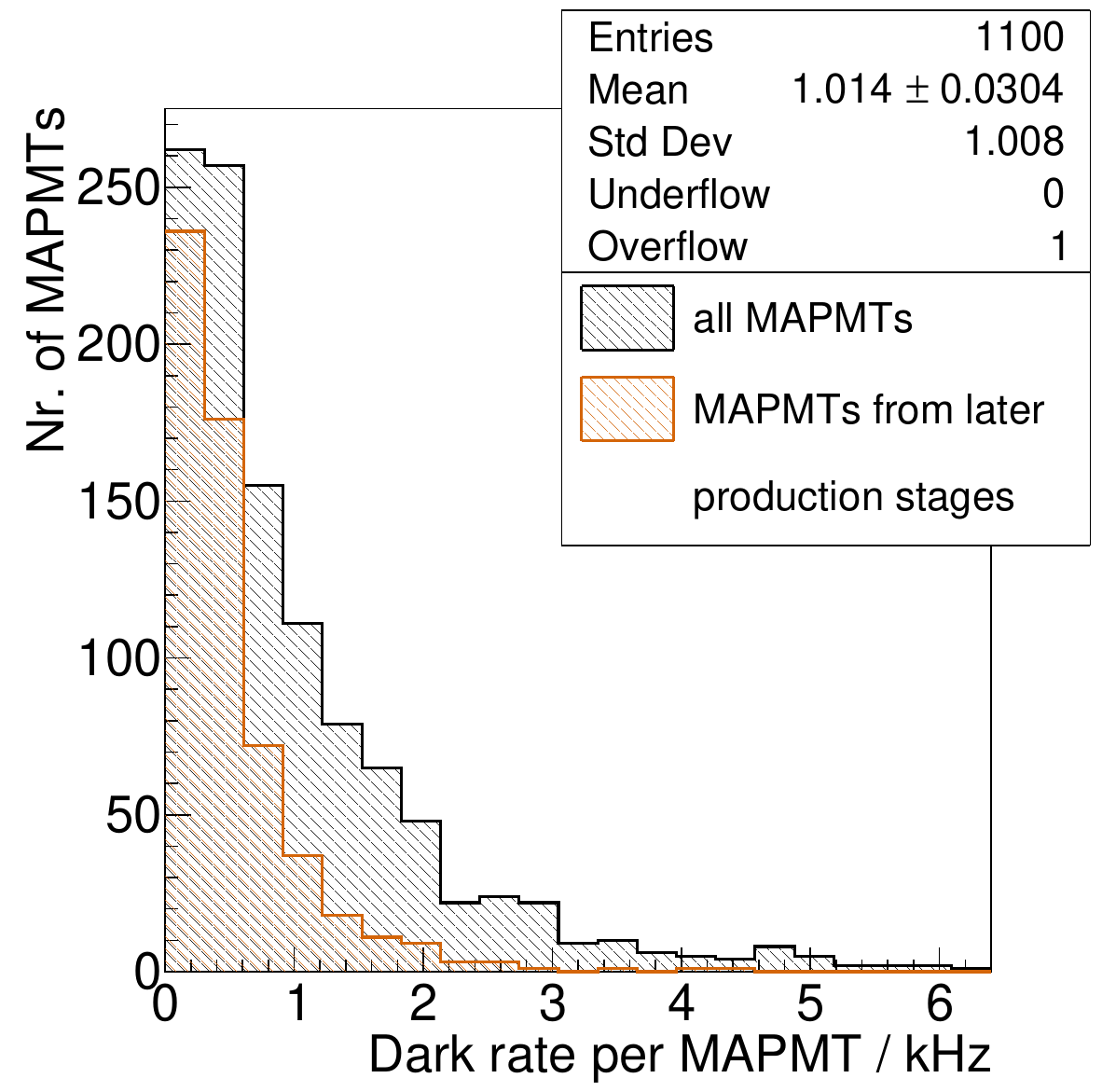}
	\caption{Integrated dark rates of each of the 1,100 H12700 MAPMTs. A subset of the full distribution is overlaid, showing only the integrated dark rates of MAPMTs produced in the second half of the production period.}
	\label{fig:dark_hist}
\end{figure}

The mean value of this distribution
$(2.64 \pm 0.25)\times 10^6$ is significantly larger than the typical gain of $1.5\times10^6$ as specified in the H12700 datasheet (cf.~\cite{Datasheet}).
In fact, fewer than 100 of the 1,100 MAPMTs have an SPE gain lower than $1.5\times10^6$ at $-1000\,$V and approximately 40~MAPMTs (3.5\%) have a gain of $4\times10^6$ or greater. 
It is worth noting that the gain stated in the individual data sheet of each MAPMT agrees, within uncertainties, with the gain measured using the test bench, despite the differing measurement techniques.
This indicates good agreement and a slightly larger measured SPE gain by the test bench in the region of interest.
The width of the observed gain distribution of $0.7\times10^6$ is significantly larger than the measurement uncertainty of the test bench. This reveals that larger variances in the average SPE gain are to be expected for this type of MAPMT.
However, these variations can be controlled by adjusting the supply voltage of each MAPMT, if a homogeneous gain distribution is important.

Figure~\ref{fig:gain_vs_HV} depicts the gain of all MAPMTs as a function of the supply voltage. Each line displays the behavior of a different MAPMT. 
Small markers and connected dashed lines indicate measurements where the center of the SPE peak moved outside the dynamic range of the ADC, in which case  the endpoint of the SPE spectrum is used to determine the gain.
The gain measured as a function of the supply voltage again visualizes the large spread in gain of all the measured MAPMTs. 
It also shows that nearly the entire range could be mitigated by operating the MAPMTs at different supply voltages. 
Following the derivation of \cite[p.~49]{HAMPhotonBook} one expects the following relation: 
\begin{equation}
    \mu = a^{10} \cdot\left(\frac{V}{10+1}\right)^{10\,k}
    \label{ham_eq}
\end{equation}
for this ten dynode MAPMT with gain $\mu$ and supply voltage $V$. 
This equation is fitted to the spectrum using the data from all MAPMTs in the $950\,$V to $1100\,$V range, where only a few measurement points are extrapolated using the SPE spectrum's endpoint.
The function and fit accurately capture the general trend of the data with $k\approx0.7$ being at the lower end of the estimate given in \cite[p.~48]{HAMPhotonBook}.

Pixel-to-pixel variations in the SPE gain are of even larger importance for the application of the MAPMTs.
Figure~\ref{fig:gain_var_hist_2d} shows the relative gain per pixel, normalized to the average gain of each MAPMT, averaged over all 1,100 measured MAPMTs.
The spatially resolved figure clearly shows a general trend of the outer rim pixels of this MAPMT model having slightly higher gain, in the order of 20\% compared to the inner pixels. However, individual MAPMTs can show larger gain variations.
Analyzing the variance of the relative gain per pixel reveals that the average over all 1,100 MAPMTs is
$(0.33 \pm 0.25)\times 10^6$.
This means that 13\% of the pixels of an average MAPMT exhibit a variation in SPE gain exceeding $0.5 \times 10^6$.

\subsection{Dark rate} 
\label{sub:dark_rate}
Another important characteristic of this MAPMT model, especially in view of the free streaming data acquisition scheme foreseen for the CBM detector, is its dark rate.
A maximum total MAPMT dark rate of $\le6.4\,$kHz, or $100\,$Hz/pixel averaged over the full MAPMT has been agreed upon with Hamamatsu for the acquisition of this particular set of 1,100 MAPMTs at an ambient temperature of $25\,^\circ$C. 

Figure~\ref{fig:dark_hist} depicts the variation of the total dark rate for all 1,100 MAPMTs, showing just a single entry above $6.4\,$kHz.
The mean total dark rate is only 
$(1.01 \pm 0.31)\,\mathrm{kHz}$, with a pronounced tail towards higher values.
This tail largely results from an increased dark rate in the early production stages, as can be seen by histogramming the integrated dark rate of those MAPMTs produced in the second half of the production period.

\begin{figure}[tb]
	\centering	\includegraphics[width=0.75\linewidth]{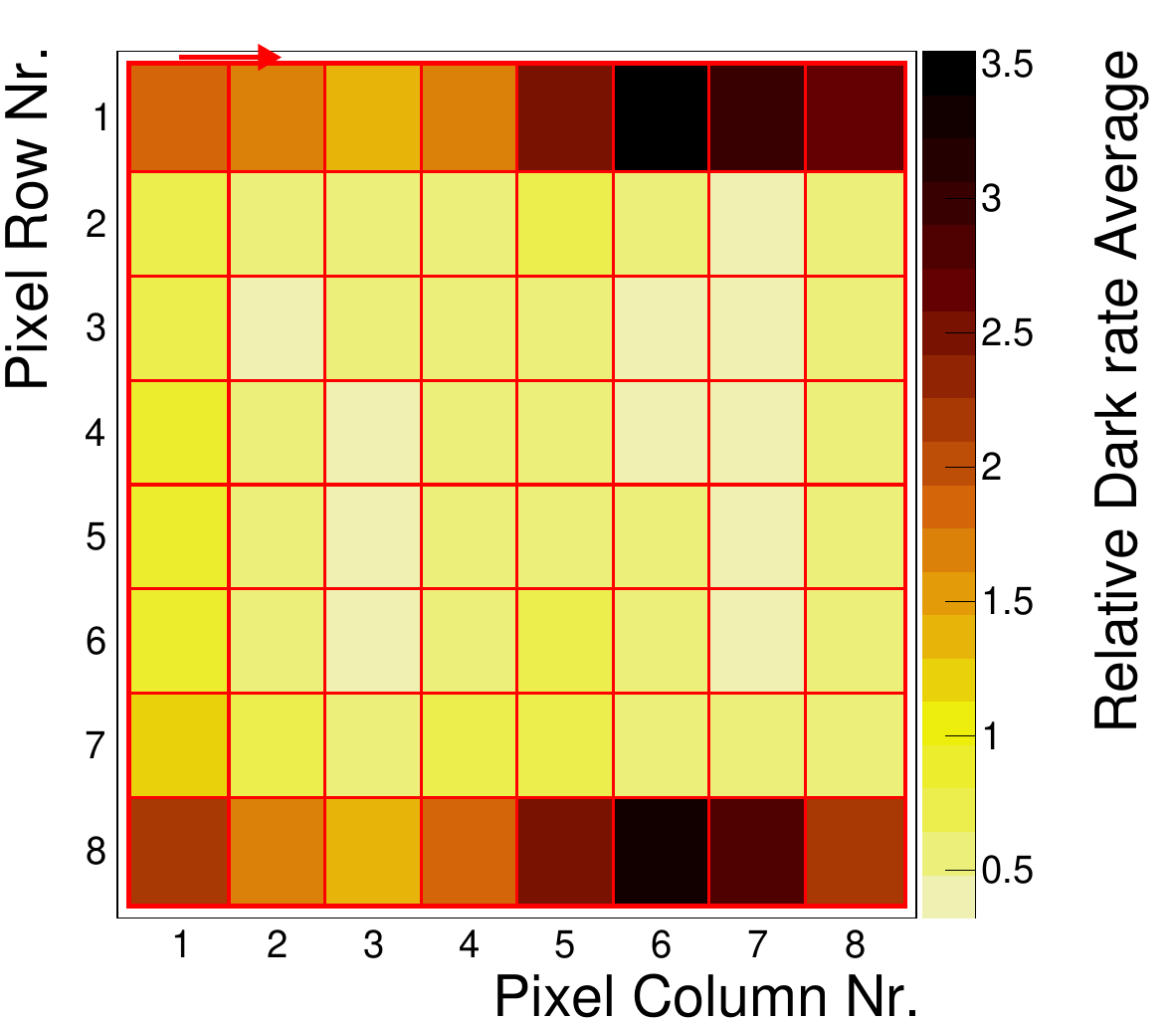}
	\caption{The dark rate per pixel is depicted in two dimensions and is averaged over all MAPMTs. It is compared to the corresponding average dark rate of the MAPMTs. The Hamamatsu ``START MARK'' is indicated by an arrow in the top left corner.}
	\label{fig:dark_var2d}
\end{figure}

Figure~\ref{fig:dark_var2d} shows the dark rate per pixel relative to the average dark rate per MAPMT as an average over all MAPMTs.
Generally, mostly the outer pixels (1 to 8 and 56 to 64) have a significantly higher dark rate than the inner pixels.
Examining individual MAPMTs reveals that single random pixels in these regions mostly contribute to the increased average.


\subsection{Efficiency} 
\label{sub:integrated_efficiency}

\begin{figure*}[hbt]
	\centering
	\includegraphics[width=\textwidth]{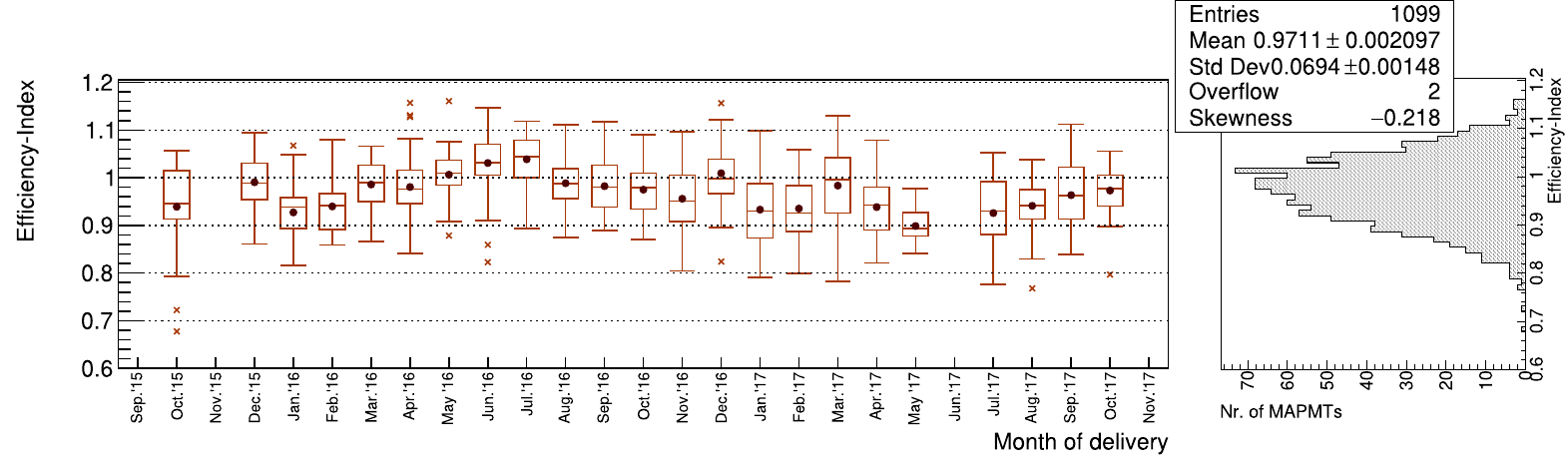}
	\caption{The box plot shows the ``Efficiency-Index'' grouped by production batch. The corresponding distribution over all MAPMTs is shown as a histogram on the right.
	The ``Efficiency-Index'' quantifies the average efficiency of each MAPMT relative to a fixed reference MAPMT.
	Each box shows the interquartile range (IQR) with the median as the center line. Whiskers extend to data within twice the IQR and outliers are marked by crosses. Dots indicate the monthly mean with systematic error bars.}
	\label{fig:eff}
\end{figure*}

\begin{figure}[tb]
	\centering
	\includegraphics[width=0.75\linewidth]{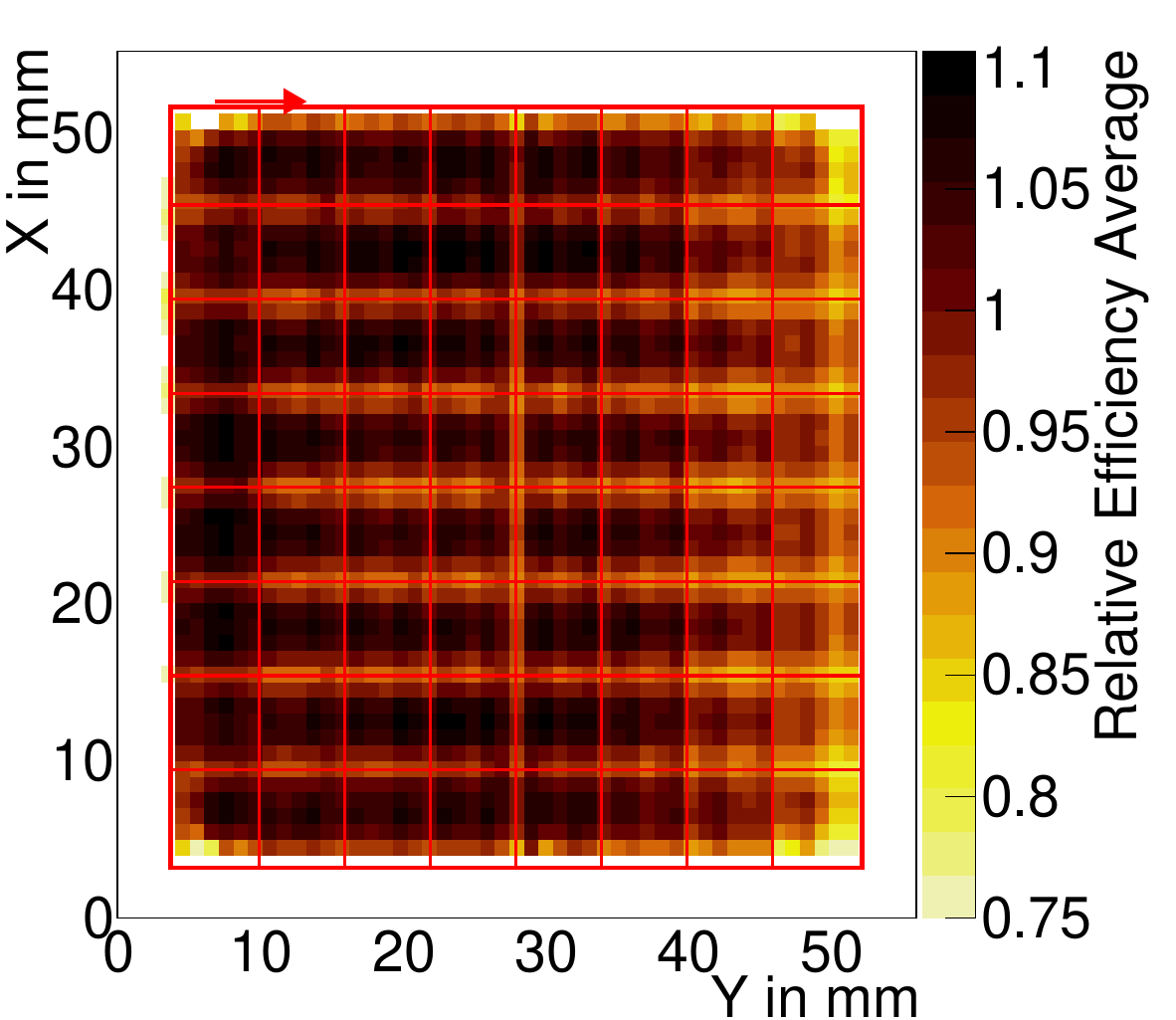}
	\caption{Average efficiency of all MAPMTs per measurement point. The Hamamatsu ``START MARK'' is indicated by an arrow on the top left corner, and the borders between the MAPMT pixels are indicated by the red boxes.}
	\label{fig:eff_var2d}
\end{figure}

Only relative efficiency values were obtained during the series of tests due to the absence of an absolute calibration of the LED light source.
As already described in Sec.~\ref{sec:the_test_bench_set_up}, the individual MAPMT efficiency is characterized relative to the reference MAPMT using an ``Efficiency-Index'', which enables the study of efficiency variations across the production period, as well as the typical variation of the spatial efficiency over the MAPMT surface.
Figure~\ref{fig:eff} depicts the development of the ``Efficiency-Index'' for all tested MAPMTs as a function of production date, together with the overall distribution of the obtained ``Efficiency-Index'' values. 
A comparison of the ``Efficiency-Index'' distributions across different production batches reveals no clear trend. While the batch averages vary, they remain within the overall spread of the full distribution. Each production batch exhibits considerable variation, resulting in an overall standard deviation of 
$(6.9 \pm 0.4)\%$ 
across all measured MAPMTs.
This indicates that fewer than $2.5\%$ of all MAPMTs have an efficiency lower than $85.2\%$, compared to the average MAPMT efficiency.

\begin{figure*}[bt]
    \centering
    \includegraphics[width=0.75\textwidth]{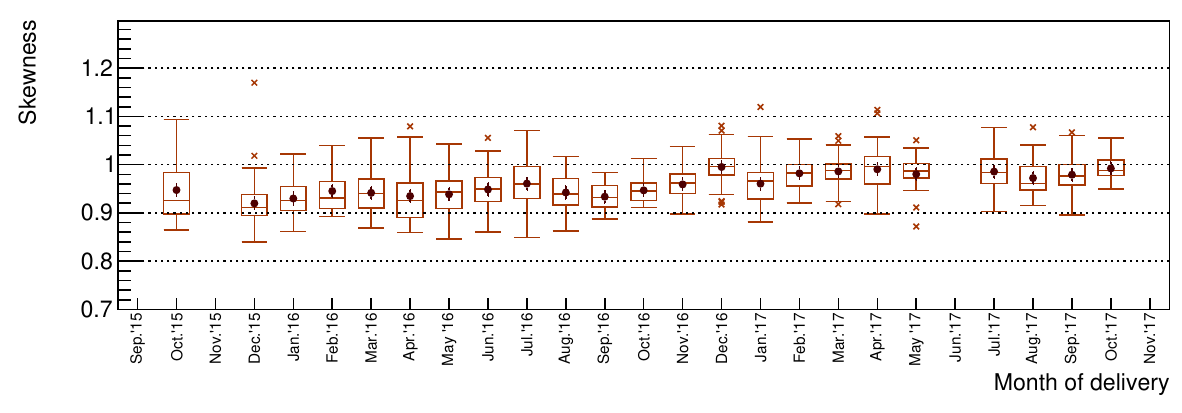}
	\caption{Box plot of the homogeneity-related skewness parameter, grouping all MAPMTs from one month. 
	Skewness is defined as the average efficiency of the right half of the MAPMT divided by the average efficiency of the left half. See Fig.~\ref{fig:eff} for details on how to read the box plot.}
	\label{fig:skew_devel}
\end{figure*}

To better understand the typical spatial patterns of the single photon detection efficiency, the relative efficiency was averaged across all 1,100 MAPMTs. The resulting spatial distribution, shown in Fig.~\ref{fig:eff_var2d}, highlights systematic variations across the MAPMT surface.
several interesting observations were made by studying this spatial efficiency. 
In the vertical direction, the individual lines of pixels are clearly separated by narrow horizontal gaps of slightly reduced efficiency. In the horizontal direction, however, there is hardly any visible separation between columns of pixels visible, except between pixel columns 4 and 5, where a clear structural separation can also be seen above the dynode grid.
This clearly reflects the inner structure of the MAPMT.
Defining an arbitrary threshold at $75\%$ of the MAPMT's average efficiency allows us to determine the active area of each individual MAPMT. This value is in good agreement with the actual cathode surface area of this MAPMT type ($48.5 \times 48.5\,\textrm{mm}^2$).
Only 33 MAPMTs in total showed active area values more than $5\,$pp.\ below the structurally constrained maximum possible effective area of $87\%$.
This good agreement between active area and cathode surface area holds true even for the altered H12700 with a decreased front window size, technically resulting in effective areas of up to $88\%$.
Finally, one recognizes an efficiency gradient that spans horizontally from higher efficiency values on the left to lower values on the right.
To assess how pronounced this effect is for each individual MAPMT, one can define a skewness parameter by dividing the average efficiency on the right half of the MAPMT by the average efficiency on the left half.
Figure~\ref{fig:skew_devel} shows the development of this skewness parameter for individual MAPMTs as a function of production batch. 
There is a clear improvement in efficiency homogeneity, which is a direct result of feedback given to the manufacturer and may be traced back to the cathode evaporation process.

\subsection{Afterpulse probability} 
\label{sub:afterpulse_probability}

\begin{figure}[tb]
	\centering
	\includegraphics[width=0.75\linewidth]{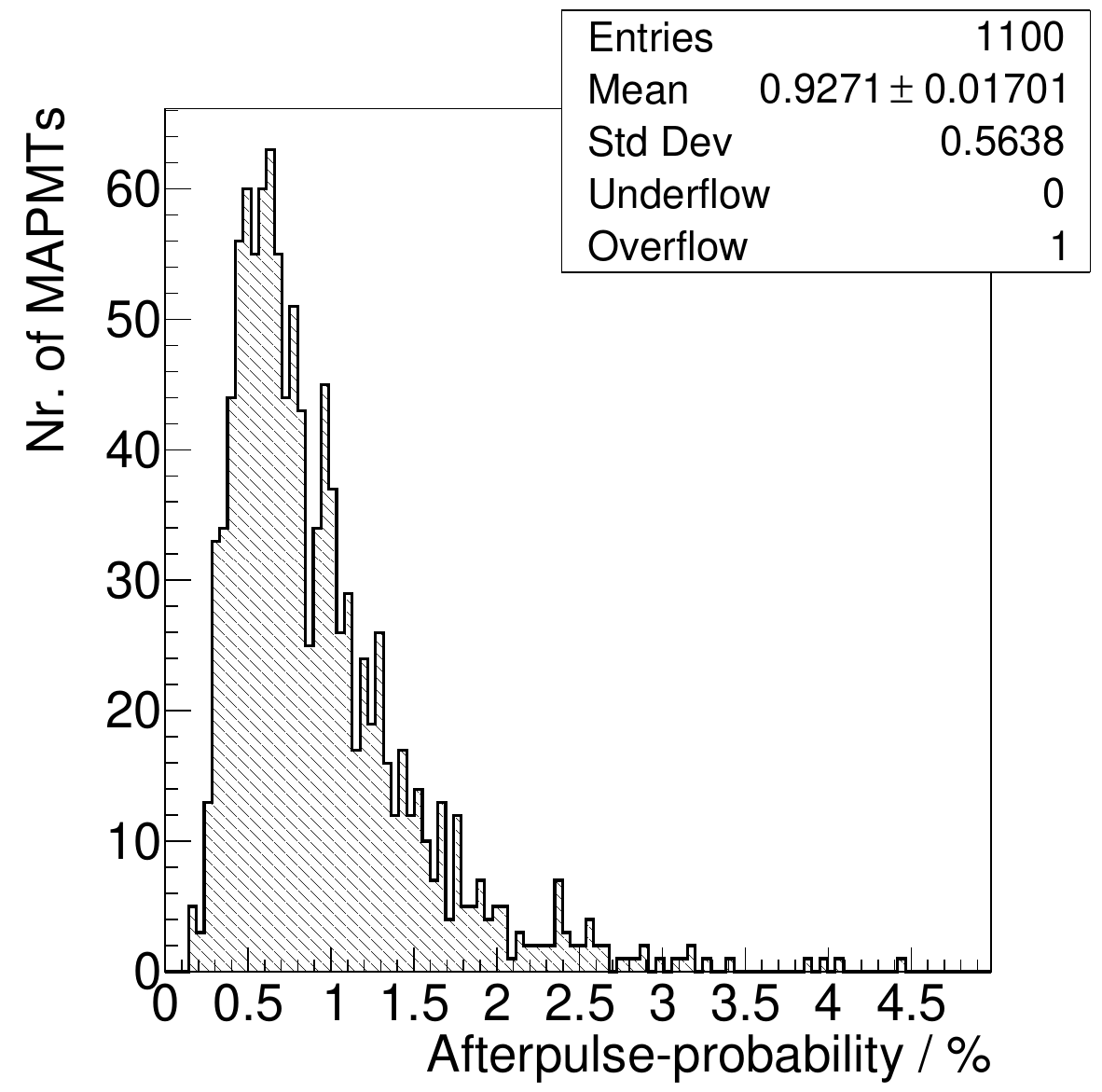}
	\caption{Average afterpulse probability for each of the 1,100 H12700 MAPMTs.}
	\label{fig:after_hist}
\end{figure}

Figure~\ref{fig:after_hist} shows the afterpulse probability, which is defined as the probability of obtaining a second single photoelectron-equivalent pulse in the same or a neighboring channel (see definition in Sec.~\ref{sec:the_test_bench_set_up}).
The probability of afterpulses following single-photon pulses in this type of MAPMT is rather low, as demonstrated by the mean value of the distribution, which is $0.93\%$ with a standard deviation of $0.56\%$.
However, it should be noted that the afterpulse probability for more recently produced MAPMTs appears to increase over time. This topic will be discussed further in Sec.~\ref{sec:long_term_variations_for_single_mapmts}.


\subsection{Notable manufacturing defects} 
\label{sub:notable_manufacturing_defects}

\begin{figure}[tb]
    \centering\includegraphics[width=0.50\linewidth]{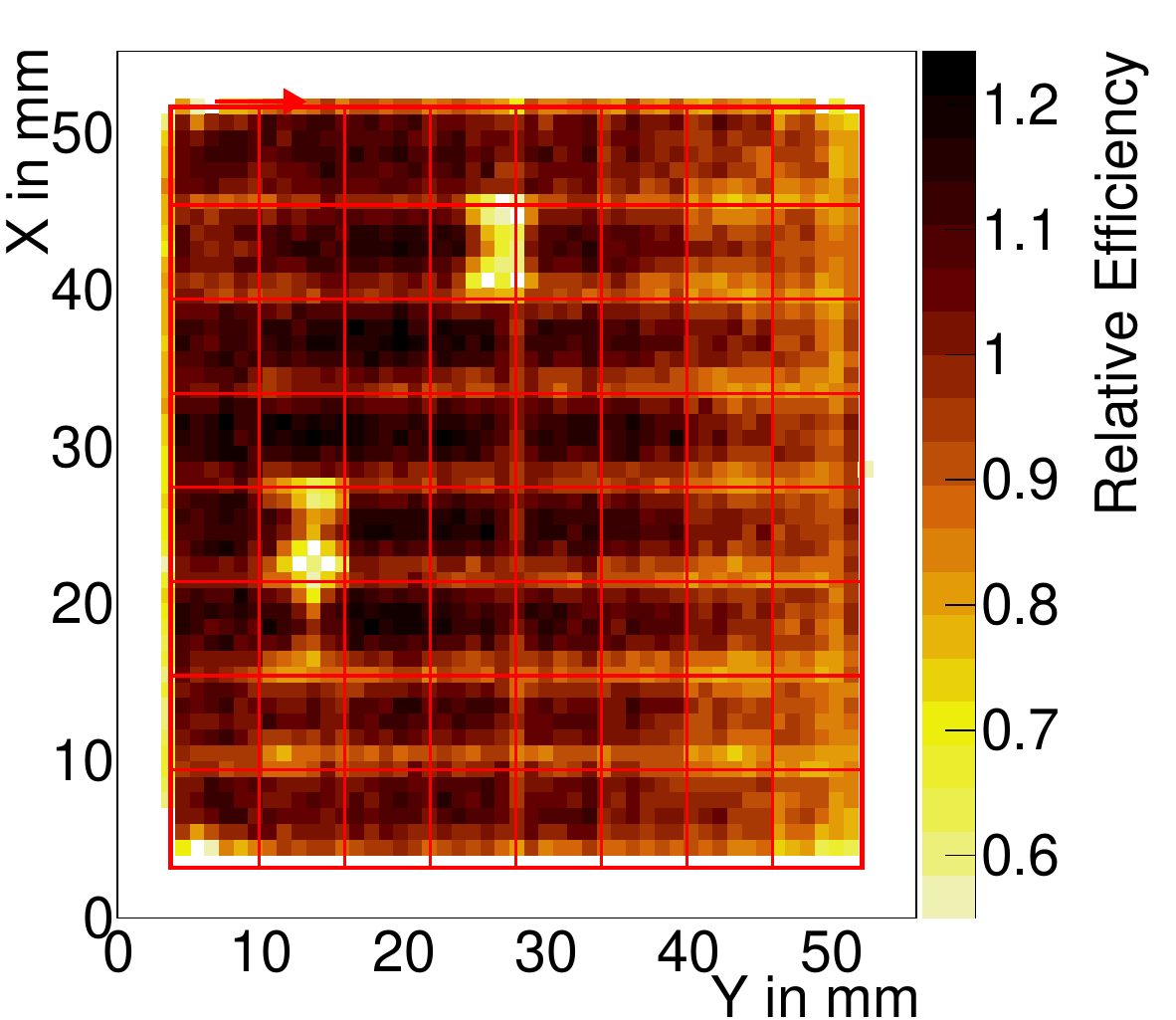}%
    \includegraphics[width=0.50\linewidth]{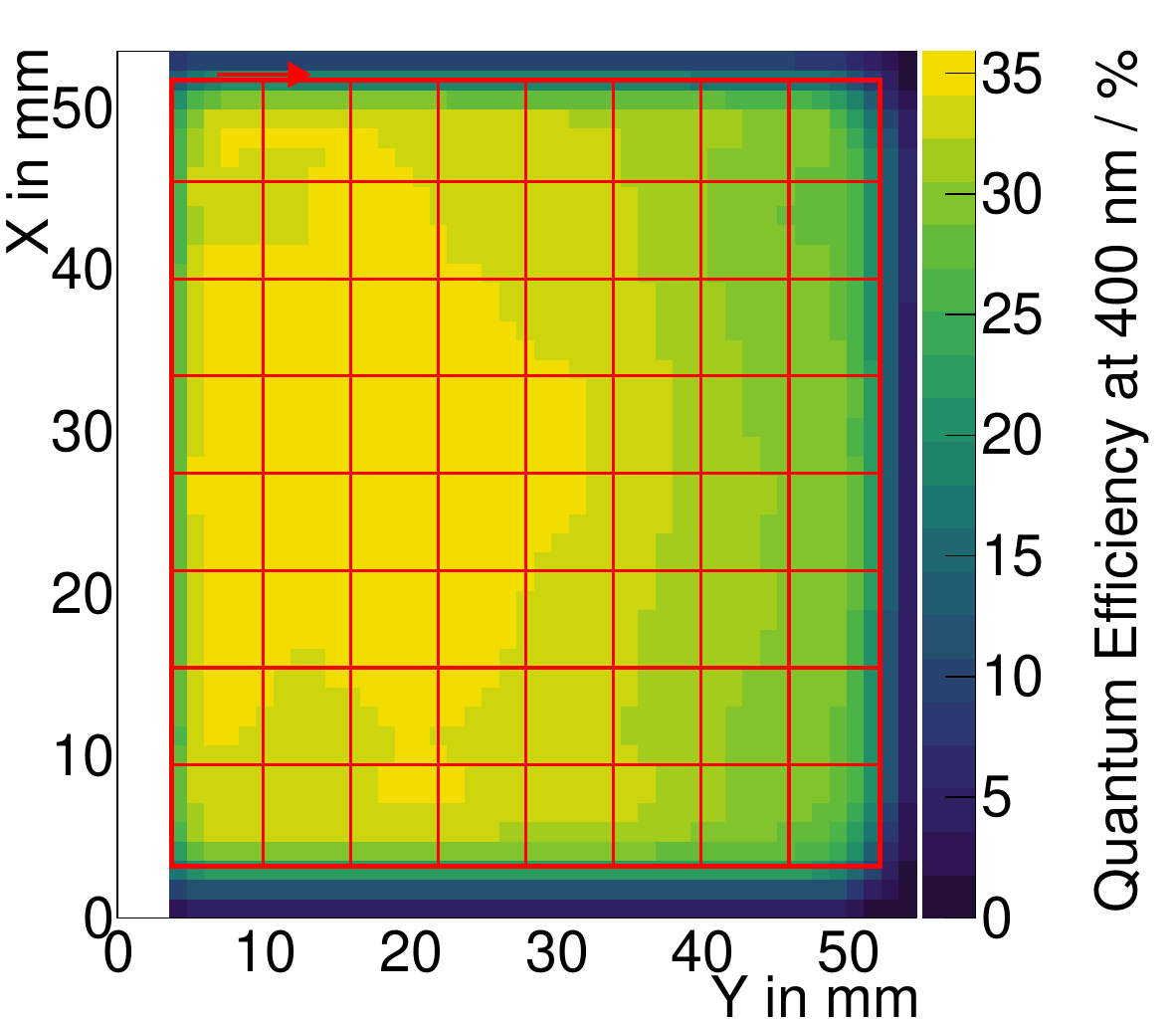}
\includegraphics[width=0.50\linewidth]{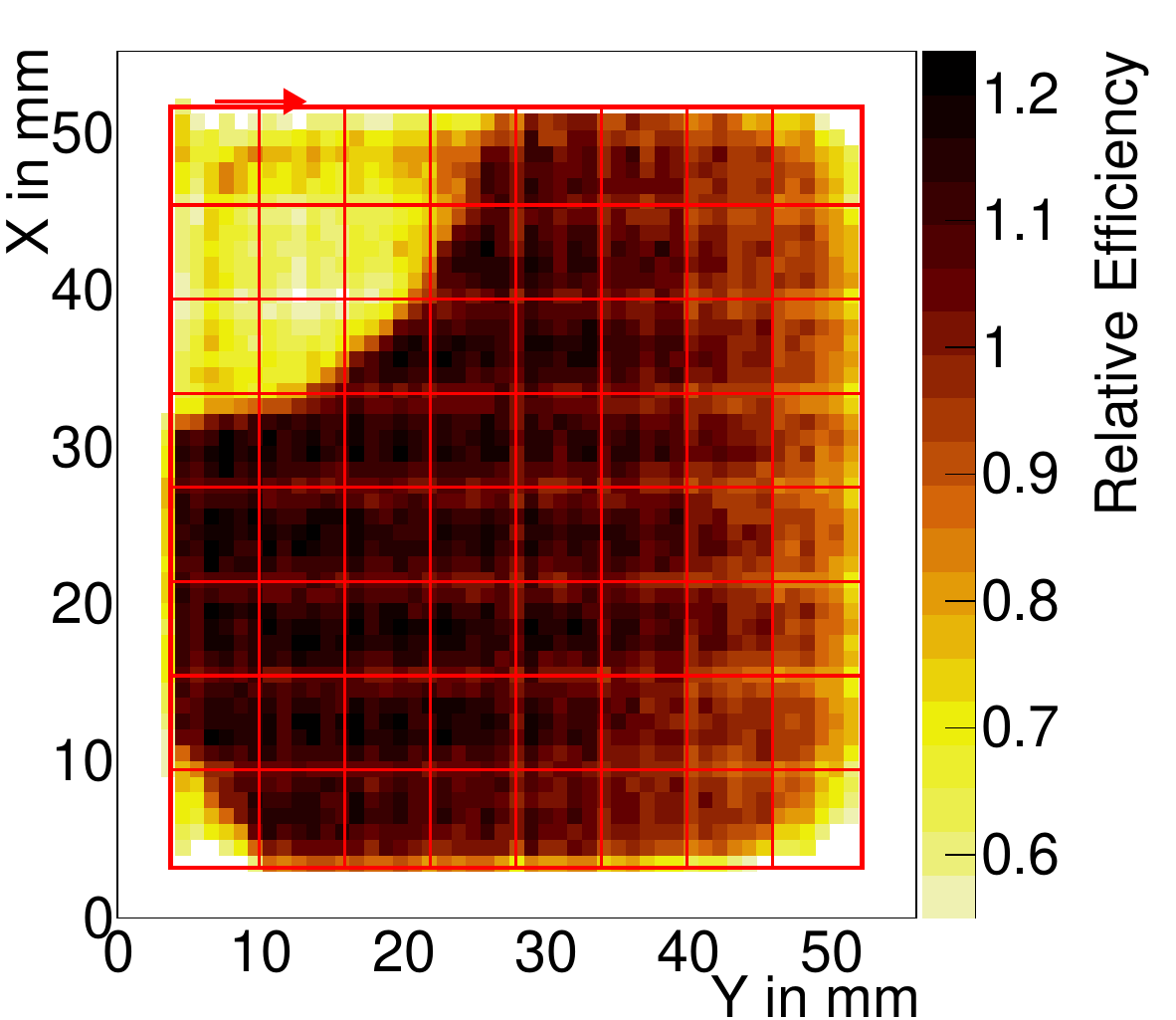}%
	\includegraphics[width=0.50\linewidth]{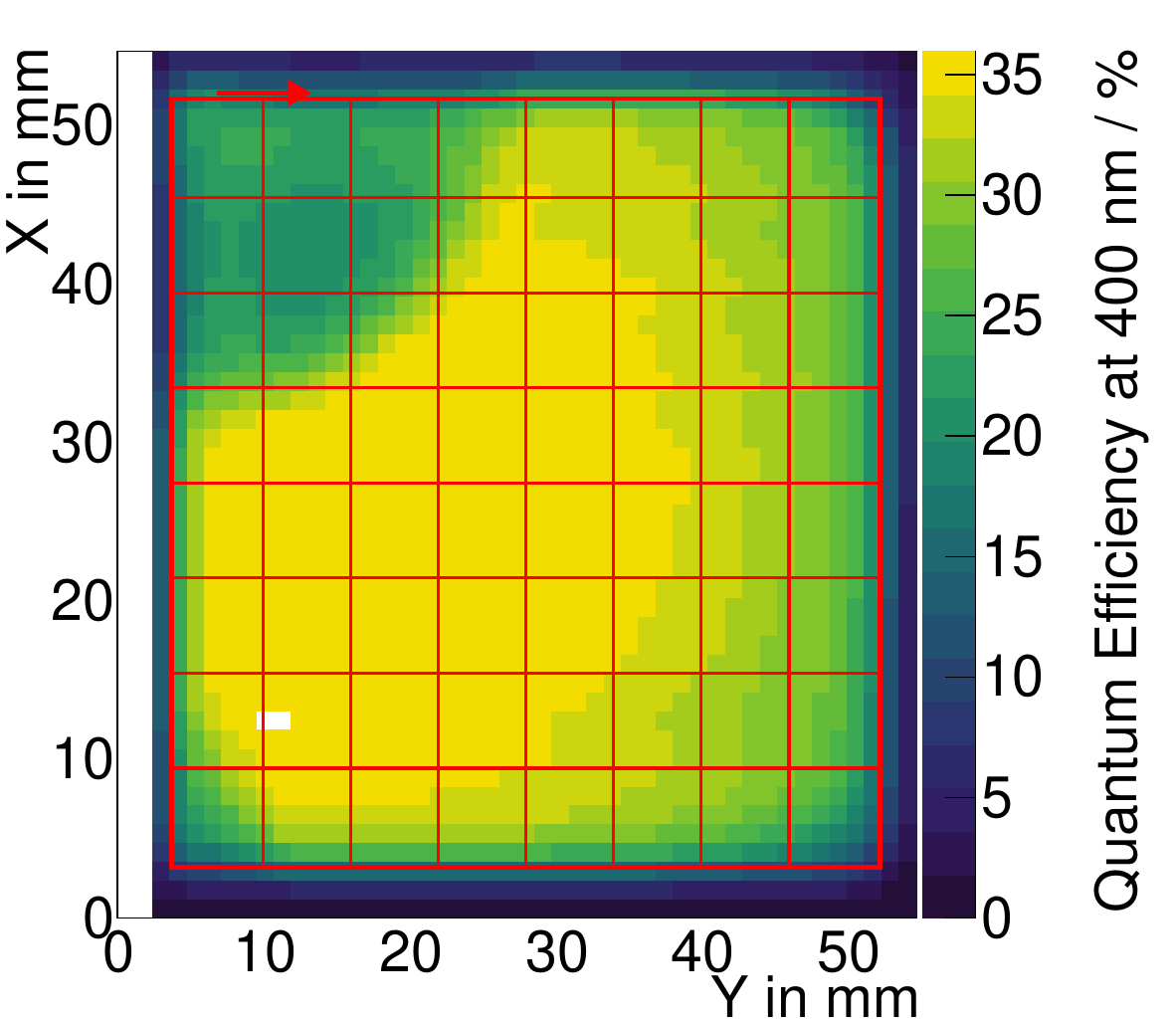}
	\caption{Spatially resolved relative single photon efficiency (left) and corresponding spatially resolved quantum efficiency measurement (right) for two single MAPMTs (top/bottom). 
	The MAPMT in the top row exhibits a deficiency due to a defect in the amplification chain or focusing mesh.
	The MAPMT in the bottom row shows a clear defect of the photocathode.
	White spots in the spatially resolved quantum efficiency scan indicate missing measurement data.}
    \label{fig:qe_spatial_single}
\end{figure}

After thoroughly testing 1,100 MAPMTs, only 19 were returned to the manufacturer for not meeting the stringent selection criteria defined in the mutually agreed specifications.
The most common reason for return was an increased dark rate, which was found in twelve MAPMTs. In these MAPMTs, mostly single pixels exhibited dark rates well above $1\,$kHz.
Other interesting defects observed in a few MAPMTs were large local inhomogeneities in the detection efficiency and ``dead spots''.
These inhomogeneities were due partly to defects in the photocathode itself and partly to flaws in the dynode system or focusing mesh.
These effects can be distinguished by measuring the spatially resolved quantum efficiency of selected, suspicious MAPMTs using a separate setup detailed in \cite{Kopfer_2014}.
Neither type of defect is visible to the naked eye. They were found in 16 of the 1,100 MAPMTs and sometimes only extended over half a pixel.
Figure~\ref{fig:qe_spatial_single} illustrates both defect types. 
It shows spatially resolved measurements of the relative single-photon detection efficiency and the spatially resolved quantum efficiency for two single MAPMTs.
The effect of a deficient photocathode is clearly visible in the bottom row, in both the quantum efficiency and the single-photon scan.
The plots in the top row depict an MAPMT with defects in a single amplification channel or small parts of the focusing mesh. In this case, inefficiencies are only visible in the single-photon scan, but not in the quantum efficiency measurement.

\section{Long-term variations for single MAPMTs} 
\label{sec:long_term_variations_for_single_mapmts}
\begin{figure*}[!hbt]
	\centering
	\includegraphics[width=0.7\textwidth]{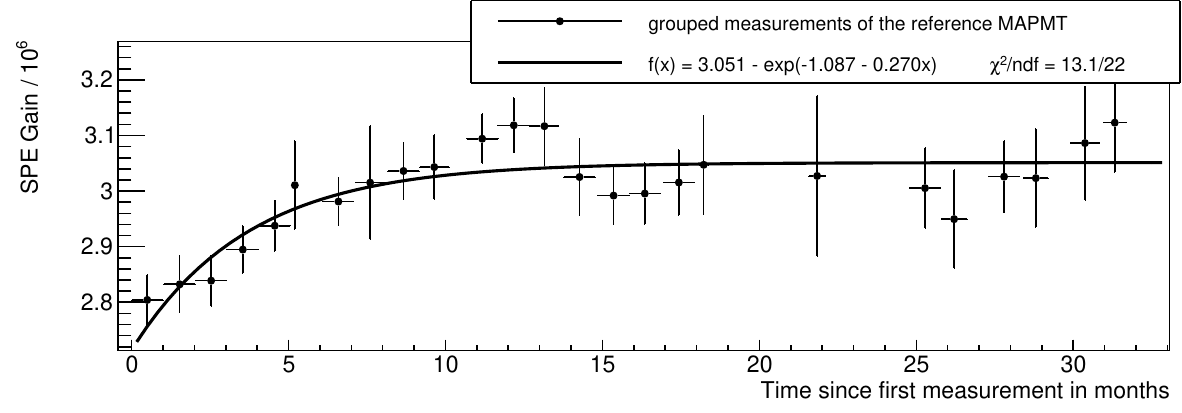}
	\caption{Gain of the main reference MAPMT as a function of the measurement date. This MAPMT has consistently been inserted into the test bench and has served as reference for efficiency measurements. The measurements are grouped by month to improve visibility.}
	\label{fig:ref_gain}
	\includegraphics[width=0.7\textwidth]{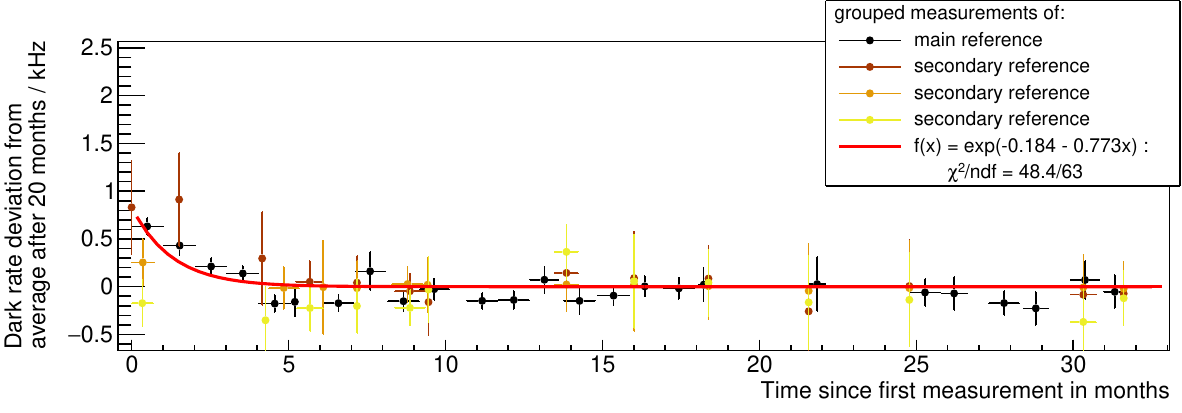}
	\caption{Absolute deviation of the dark rate (in kHz) relative to the rate measured 20 months after delivery, shown as a function of time for the main and secondary reference MAPMTs. Data from each month are grouped together to improve visibility. All points are fitted with an exponential decay curve.}
	\label{fig:ref_dark}
	\includegraphics[width=0.7\textwidth]{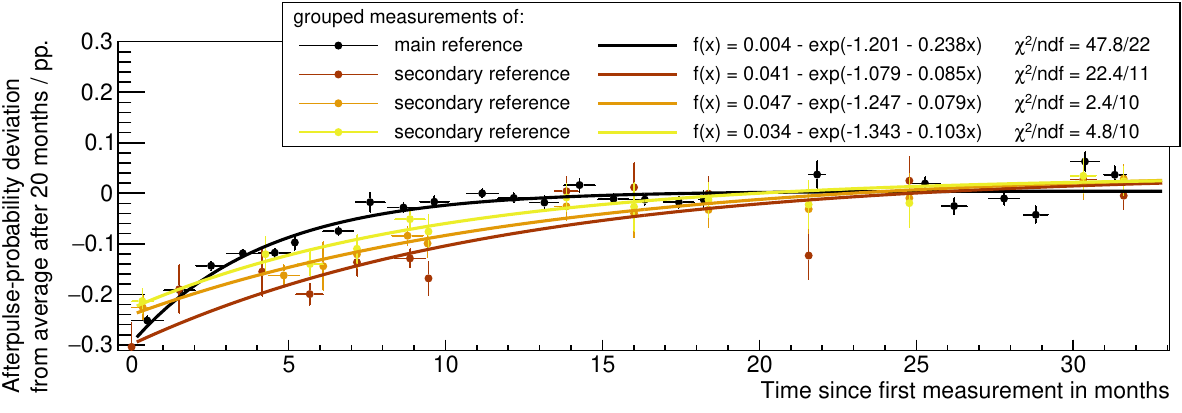}
	\caption{Afterpulse probability of the main and secondary reference MAPMTs. The visualization follows Fig.~\ref{fig:ref_dark}. Each individual MAPMT is fitted with an exponential increase.}
	\label{fig:ref_after}
\end{figure*}

The primary reference MAPMT remained in the test bench for most of the three-year measurement period. As such, it was operated and scanned repeatedly during approximately 400 individual measurements, each lasting about eight hours. Over the full time period, three additional secondary reference MAPMTs were scanned repeatedly together with the primary reference MAPMT. These secondary reference MAPMTs were stored under dark conditions in a nitrogen-flushed compartment and served as a cross-check for the primary reference MAPMT.
Repeated measurements of these reference MAPMTs enable the study of long-term variations in certain absolute quantities, such as gain, dark rate, and afterpulsing, which are independent of normalization. This section describes such analyses and their results.
Figure~\ref{fig:ref_gain} shows the SPE gain (averaged over all pixels) of the reference MAPMT over time. Measurements are grouped monthly to improve visibility by reducing variations from measurement uncertainties.
The data points are well described by an exponential increase that flattens out after approximately ten months of operation.
The systematic increase in gain corroborates the findings in \cite{Calvi:2015yra}, which describe long-term variations in the gain for this type of MAPMT after extended periods of illumination.
In this setup, however, the illumination intensity of the MAPMT is very low, because only single photons are emitted onto small spots on the MAPMT. Therefore, it is reasonable to assume that high voltage operation alone, even without illumination, might cause variations in the gain.

Figure~\ref{fig:ref_dark} shows the long-term development of the dark rate for all four MAPMTs.
An exponential-like initial decrease in the dark rate is observed for the reference MAPMT and two other repeatedly measured (secondary reference) MAPMTs. A few months after delivery, all MAPMTs reach a relatively stable plateau.
These observations suggest that new MAPMTs of this type should be expected to settle over the course of a few months, but no further increase should be expected without additional illumination or irradiation. The time and amount of this settling process may vary from MAPMT to MAPMT.

A similar, yet opposite trend is visible when measuring the afterpulse probability of the same four MAPMTs, as shown in Fig.~\ref{fig:ref_after}.
All four MAPMTs exhibit a similar increasing trend, starting at an afterpulse probability of approximately $0.3\,$pp.\ lower and reaching their final values over varying time spans.
Afterpulsing is commonly attributed to residual gas and helium diffusing into the MAPMT.
Notably, all the secondary reference MAPMTs were stored in a nitrogen-flushed compartment for most of the time between measurements, while the main reference MAPMT remained in the setup under ambient atmospheric conditions.



\section{Conclusion} 
\label{sec:conclusion}
This article presents the results of a series of tests and characterization measurements conducted on 1,100 H12700 MAPMTs delivered by Hamamatsu.
The tests were performed using a dedicated test bench. Each MAPMT underwent $x$-$y$ scanning of the photocathode with pulsed single photons, and the responses of all channels were recorded in parallel using a time- and charge-sensitive, self-triggered data acquisition system. This setup allows the extraction of several key performance characteristics to be extracted from a single data set.
Key observations were made by measuring the gain, dark rate, efficiency, homogeneity and afterpulse probability. The average single photoelectron gain was for most MAPMTs higher than the typical gain stated by Hamamatsu.
The pixels with the largest gains are mostly found on the outer rim of the MAPMT.
Furthermore, there is a large overall spread in gain between the different MAPMTs, which can be counteracted by adjusting the high voltage.
The top and bottom rows (pixels 1--8 and 56--64) have the highest dark rates, with an average dark rate per full MAPMT of approximately $1\,$kHz.
Overall, the dark rate decreased with operation time for each MAPMT but was also lower for later production batches.
For the initial production batches of the H12700, a left-right inhomogeneity of the photocathode was discovered. This feature was less pronounced for later production batches.
The afterpulse probability was found to initially increase with operation and storage time, reaching its maximum a few months after production. The average afterpulse probability across all studied MAPMTs is slightly below $1\%$.
Further analyses revealed that large efficiency defects in individual MAPMTs are rare and are usually caused by flaws of the photocathode, localized imperfections in the single amplification chains, or defects in the focusing mesh.
Of all the 1,100 MAPMTs delivered, more than $98\%$ directly met the strict selection criteria agreed upon with the manufacturer, particularly with regard to single-pixel and overall dark rates. Based on these measurement results, a few MAPMTs were replaced.

\section*{Acknowledgments} 
\label{sec:acknowledgements}
This work was supported by GSI, by BMBF Grants 05P15PXFCA, 05P19PXFCA, 05P21PXFC1 and 05P24PX1, and by ``Netzwerke 2021'', an initiative of the Ministry of Culture and Science of the State of Northrhine Westphalia.


\bibliography{main}


\end{document}